\documentclass[12pt]{iopart}

\usepackage{iopams}
\usepackage{graphicx}
\usepackage{subfigure}
\usepackage{mcite}
\usepackage{hyperref}

\begin{document}

\title{Competition between superconductivity and nematic order in high-$T_c$ superconductor}

\author{Jing Wang$^1$ and Guo-Zhu Liu$^{1,2}$\footnote[7]{Author to whom any correspondence
should be addressed.}}

\address{$^1$Department of Modern Physics, University of Science and
Technology of China, Hefei, Anhui, 230026, P.R. China}
\address{$^2$Max Planck Institut f$\ddot{u}$r Physik komplexer
Systeme, D-01187 Dresden, Germany}
\ead{gzliu@ustc.edu.cn}

\begin{abstract}
We investigate the competition between $d$-wave superconductivity
and nematic order in high-$T_c$ superconductor, and examine the role
played by gapless fermionic degrees of freedom. Apart from the
competitive interaction with the superconducting order parameter,
the nematic order parameter couples strongly to gapless nodal
quasiparticles. The interplay of these two kinds of interactions is
analyzed by means of renormalization group method. In case the
fermionic degree of freedom are entirely neglected, the competitive
interaction between two bosonic order parameters is strongly
relevant, and can lead to runaway behavior. However, these
properties are fundamentally changed once the dynamics of fermions
are taken into account. At the nematic quantum critical point where
an extreme fermion velocity anisotropy occurs, the superconducting
and nematic order parameters are decoupled from each other.
Consequently, the phase transitions are continuous, and $d$-wave
superconductivity can coexist with nematic order homogeneously.
These results indicate that the gapless fermions can play an
important role and should be carefully included in the theoretical
description of competing orders.
\end{abstract}


\maketitle

%%%%%%%%%%%%%%%%%%%%%%%%%%%%%Main Body%%%%%%%%%%%%%%%%%%%%%%%%%%%%%%%%%%%%%

\section{Introduction}

Unconventional superconductors usually refer to the superconductors
those can not be understood within the conventional
Bardeen-Cooper-Schrieffer (BCS) theory. Notable examples of
unconventional superconductors are high-$T_c$ cuprate
superconductor, heavy fermion superconductor, and iron-based
superconductor. Unlike BCS superconductors, unconventional
superconductivity is generally driven by electron-electron
interactions, and often has a magnetic origin. Another interesting
property of unconventional superconductor is that its ground state
is not unique. In addition to the defining superconducting state,
unconventional superconductors also exhibit a variety of other
symmetry-broken ground states, including antiferromagnetc, nematic,
and stripe states, upon tuning such parameters as doping and
pressure \cite{Kivelson03, Vojta, Fradkin, heavyfermion, Flouquet}.
A widely recognized notion is that the long-range superconducting
order competes, and under certain circumstances coexists, with other
long-range orders. The competition and possible coexistence between
different orders can give rise to rich properties, and hence have
attracted intense theoretical and experimental interest in the past
years.

The successful microscopic theory of competing orders has not yet
been established to date, primarily because the pairing mechanism in
most unconventional superconductors is still undetermined. A
realistic and commonly used strategy is to build low-energy
effective field theory on phenomenological grounds. One can first
write down the Ginzburg-Landau (GL) actions for two bosonic order
parameters and then introduce certain coupling terms between these
two scalar fields. Such generalized GL model has recently been
applied to describe competing orders in a number of unconventional
superconductors \cite{Arovas, Kivelson02, Demler, Moon, Zaanen,
Nussinov, Millis10, Chubukov, Chowdhury, Schmalian}. An early
success of such theoretical investigation is the prediction of
field-induced antiferromagnetic core in the supercondcuting vortices
of high-$T_c$ superconductors \cite{Arovas}. This prediction was
subsequently confirmed in experiments \cite{Lake1, Lake2}.
Interestingly, experiments further found that the antiferromagnetic
order not only exists in the vortex cores, but also extends into the
superconducting region \cite{Lake1} and exhibits nontrivial spatial
modulation \cite{Lake2, Hoffman}. A phenomenological field theory
that contains a simple quadratic-quadratic coupling term between
superconducting and antiferromagnetic order parameters was put
forward to understand these new findings \cite{Demler, Kivelson02}.

Recently, the issue of competing orders has attracted revived
interest. It is found that the competitive interaction between
distinct orders can drive an instability, which gives rise to a
general tendency of first order transition \cite{Zaanen, Millis10}.
This phenomenon may account for the first order transition observed
in some unconventional superconductors \cite{heavyfermion}. In
addition, nonuniform glassy electronic phases and Brazovskii type
transitions are predicted to emerge due to competition between two
long-range orders \cite{Nussinov}. Another interesting observation
is that the competition between superconducting and
antiferromagnetic orders can help to judge the gap symmetry of
iron-based superconductors \cite{Chubukov, Schmalian}. Furthermore,
the competition between superconducting and nematic orders might be
responsible for \cite{Chowdhury} the electronic anisotropy observed
in the vortex state of FeSe superconductor \cite{Xue}.

The effective field theory adopted in previous analysis of competing
orders normally contains only two bosonic order parameters. The
fermionic degrees of freedom are usually completely integrated out
in the spirit of Hertz-Millis-Moriya (HMM) theory \cite{Hertz,
Millis, Moriya}. This integration procedure is expected to be
applicable in systems that do not contain gapless fermionic
excitations. For instance, the iron-based superconductors seem to
have a $s$-wave energy gap, so the electronic excitations are fully
gapped and can be safely integrated out \cite{Chubukov, Schmalian}.
However, such integration manipulation is not always valid. Indeed,
its validity has recently been questioned in several itinerant
electron systems \cite{Belitz, Abanov, Rech, Metzner}. In the
systems that exhibit gapless fermionic excitations, integrating out
fermions may lead to singularities, especially in the vicinity of
quantum critical point (QCP). Actually, infrared singularities have
been found on the border of several quantum phase transitions
\cite{Belitz, Rech, Abanov, Metzner}. In order to properly describe
the quantum critical behavior in these systems, it is more
appropriate to maintain both bosonic order parameter and gapless
fermions in the effective theory. When a long-range order competing
with superconductivity also couples to gapless fermions, it would be
interesting to go beyond HMM theory and examine the role of gapless
fermions. Recent analysis presented in Refs.~\cite{Moon, Liu} did
suggest nontrivial roles played by gapless fermions.

\begin{figure}
   \includegraphics[width=3.5in]{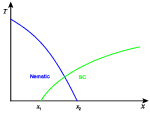}
   \centering
   \vspace{-0.35cm}
\caption{Schematic phase diagram on $(x,T)$-plane of high-$T_c$
superconductors. $x$ represents doping concentration. $x_1$ and
$x_2$ are QCPs of superconducting and nematic phase transitions,
repectively.}\label{Fig_coexist}
\end{figure}

In various types of superconductors, superconductivity may compete
with several possible long-range orders. To examine the role of
fermions, we wish to study a prototypical model which describes
competition between two distinct long-range orders, contains gapless
fermions, and in the meantime is technically controllable. In the
present paper, we choose to consider the competition between
superconductivity and nematic order in the contexts of high-$T_c$
superconductors. In recent years, there has been increasing
experimental evidence pointing towards the existence of an
electronic nematic phase in some high-$T_c$ superconductors
\cite{Kivelson03, Fradkin, Vojta, Ando, Hinkov, Daou, Lawler},
especially YBa$_2$Cu$_3$O$_{6+\delta}$ and
Bi$_2$Sr$_2$CaCu$_2$O$_{8+\delta}$. According to these experiments,
a nematic order is predicted to compete and coexist with
superconductivity, which is schematically plotted in
Fig.~\ref{Fig_coexist}. The nematic transition and the coupling of
nematic order to gapless fermions have stimulated intense research
effort \cite{Kivelson03, Fradkin, Vojta, Kivelson, VojtaSachdev,Sachdevbook,
Metznernematic, Oganesyan, Kim, Huh, Xu, Fritz, Lawler2, WLK, Liu}.
From a field-theoretic viewpoint, the nematic order parameter is a
real scalar field and does not carry a finite wave vector, which
substantially simplifies theoretical calculations.

It is known that high-$T_c$ superconductor has a $d_{x^2 - y^2}$
energy gap, which vanishes at four nodes, $\left(\pm\frac{\pi}{2},
\pm\frac{\pi}{2}\right)$. Therefore, gapless nodal quasiparticles
(qps) are present even at the lowest energy in the superconducting
phase. These nodal qps are believed to be responsible for many
anomalous low-temperature properties of the superconducting dome.
When a nematic QCP exists somewhere in the superconducting dome, as
shown in Fig.~\ref{Fig_coexist}, the fluctuation of nematic order
parameter couples to gapless nodal qps. This coupling can generate
non-Fermi liquid behaviors and other anomalous phenomena in the
vicinity of nematic QCP \cite{Kim, Huh, Fritz, Xu, Lawler2, WLK,
Liu}. In particular, the ratio $\kappa$ between gap velocity
$v_{\Delta}$ and Fermi velocity $v_F$ of nodal qps is driven to
vanish, i.e., $\kappa = v_{\Delta}/v_F \rightarrow 0$, by the
critical nematic fluctuation \cite{Huh}, leading to extreme velocity
anisotropy. These unusual behaviors may have significant effects on
the interplay between superconductivity and nematic order, which is
the topic of this paper.

In this paper, we first write down an effective field theory that
describes both the competitive interaction between two bosonic
(superconducting and nematic) order parameters and the Yukawa-type
coupling between nematic order parameter and nodal qps. We then
carry out a detailed renormalization group (RG) analysis
\cite{Shankar94} within this effective theory. Specifically, we will
derive and solve the RG flow equations of all physical parameters so
as to determine the possible stable fixed points. We will
demonstrate that gapless fermionic degrees of freedom can
fundamentally change the basic properties of the interplay between
superconductivity and nematic order. In case all the fermions are
entirely neglected, the ordering competition may be strong enough to
produce runaway behavior and turn continuous phase transitions to
first order \cite{Zaanen}. However, once the dynamics of gapless nodal
qps are properly incorporated, a stable fixed point exists with the two
originally competing bosonic order parameters decoupled from each
other in the vicinity of nematic QCP. As a result, both the
superconducting and nematic transitions remain continuous. In
addition, the \emph{d}-wave superconductivity can coexist with the
nematic order homogeneously. Our results indicate that it is
important to include the dynamics of gapless fermions in the
theoretic description of competing orders.

In Sec. \ref{sec_model}, we write down the effective action which
contains two bosonic order parameters and gapless nodal qps. In Sec.
\ref{sec_RG_analysis}, we make RG calculations and derive the flow
equations for all parameters in the effective action. In Sec.
\ref{sec_numerical_discussion}, we present numerical solutions of
the flow equations and discuss the physical implications. The paper
is ended in Sec. \ref{sec_summary} with summary and conclusion.

\section{Effective field theory of competing orders}\label{sec_model}

We first need to write down an effective field theory to describe
the competition between superconducting and nematic orders. This
will be done largely on phenomenological grounds. In the phase
diagram presented in Fig.~\ref{Fig_coexist}, the horizonal axe is
doping concentration $x$. The QCP of superconducting transition is
$x_1$, which is roughly $x_1 \approx 0.05$ in many high-$T_c$
superconductors. The anticipated QCP for nematic transition is
represented by $x_2$. So far, the precise value, and even the very
existence, of $x_2$ have not yet been unambiguously determined.
Here, we assume that $x_2$ is larger than $x_1$, which implies a
bulk coexistence of superconducting and nematic orders.

In the present system, there are three types of degrees of freedom:
superconducting order parameter $\psi$, nematic order parameter
$\phi$, and gapless nodal qps $\Psi$. The competition between
superconducting and nematic orders can be described by a repulsive
quadratic-quadratic coupling term, $\propto \psi^2 \phi^2$, which is
widely adopted in the description of competing orders
\cite{Schmalian, Moon, Zaanen, Liu}. In addition to this competitive
interaction, the nematic order parameter $\phi$ also interacts with
gapless nodal qps $\Psi$, which is usually described by a
Yukawa-type coupling term. There is, however, no direct coupling
between the superconducting order parameter and nodal qps. First,
the nodal qps are excited from the $d_{x^2 - y^2}$ gap nodes where
superconducting order parameter vanishes. Second, these qps are
known to have a sharp peak and a very long lifetime in the
superconducting dome in the absence of competing orders
\cite{Orenstein00}, so their coupling to $\psi$ must be quite weak.
Furthermore, in this paper we are mainly interested in the physical
properties in the close vicinity of nematic QCP $x_2$, where the
fluctuation of nematic order parameter is critical. Unless the
superconducting QCP $x_1$ coincides with, or is very close to,
$x_2$, the fluctuation of superconducting order parameter is not
critical at $x_2$. Therefore, the coupling between superconducting
order parameter and nodal qps is not as important as that between
critical nematic fluctuation and nodal qps at $x_2$, and can be
simply neglected.

\begin{figure}
   \includegraphics[width=4.5in]{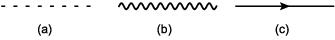}
   \centering
   \vspace{-0.4cm}
\caption{(a): Free propagator of superconducting field $\psi$; (b):
Free propagator of nematic field $\phi$; (c): Free propagator of
nodal qps $\Psi$.}\label{propagators}
\end{figure}

On the basis of the above qualitative analysis, we can write down
the following partition function
\begin{eqnarray}
Z&=&\int  \mathcal{D}\psi\mathcal{D}\phi
\mathcal{D}\bar{\Psi} \mathcal{D}\Psi e^{S},
\end{eqnarray}
where the effective action is
\begin{eqnarray}
S &=& S_{\psi} + S_{\phi} + S_{\Psi} + S_{\psi\phi} +
S_{\Psi\phi},\label{Eq_action_total} \\
S_{\psi} &=& \frac{1}{2}\int \frac{d^3q}{(2\pi)^3}(-2\alpha+q^2)
\psi^2 + \frac{\beta}{2} \int d^2\mathbf{r}d\tau\psi^{4}, \label{Eq_S_psi}\\
S_{\phi} &=& \frac{1}{2}\int \frac{d^3q}{(2\pi)^3}\left(-2r + q^2
\right)\phi^2 + \frac{u}{2} \int d^2\mathbf{r}d\tau\phi^{4},\label{Eq_S_phi}\\
S_{\Psi} &=& \int \frac{d^3k}{(2\pi)^3}
[\Psi^{\dagger}_{1i}(-i\omega +
v_{F}k_{x}\tau^{z} + v_{\Delta}k_{y}\tau^x)\Psi_{1i} \nonumber \\
&& + \Psi^{\dagger}_{2i}(-i\omega
+ v_{F}k_{y}\tau^{z} + v_{\Delta}k_{x}\tau^{x})\Psi_{2i}], \\
S_{\psi\phi} &=& \gamma \int d^2xd\tau \psi^2 \phi^2,\\
S_{\Psi\phi} &=& \int d^2xd\tau [\lambda \phi(\Psi^{\dagger}_{1 i}
\tau^{x}\Psi_{1 i} + \Psi^{\dagger}_{2i}
\tau^{x}\Psi_{2i})],\label{Eq_S_Psi_phi}
\end{eqnarray}
where $\tau^{x,y,z}$ are Pauli matrices and the flavor index $i$
sums up $1$ to $N$. $\Psi^{\dagger}_{1}$ represents nodal QPs
excited from $(\frac{\pi}{2},\frac{\pi}{2})$ and
$(-\frac{\pi}{2},-\frac{\pi}{2})$ points, and $\Psi^{\dagger}_{2}$
the other two. The physical flavor of nodal qps, $N = 2$. Here, $r$
is tuning parameter for nematic transition with $r = 0$ at $x_2$.
$v_{F,\Delta}$ are the Fermi velocity and gap velocity of nodal qps,
respectively. The competitive interaction term $\psi^2 \phi^2$ has a
positive coefficient, $\gamma > 0$. $S_{\Psi\phi}$ represents the
Yukawa coupling between nematic order parameter and gapless nodal
qps, with $\lambda$ being its coupling constant. The free
propagators of all the fields are shown in Fig.~\ref{propagators}.

In order to simplify calculations, it proves convenient to make two
transformations \cite{Huh}: $\phi \rightarrow \phi/\lambda$, and
$r\rightarrow \lambda^2 r$. It is now easy to rewrite Eq.
(\ref{Eq_S_Psi_phi}) as
\begin{eqnarray}
S_{\Psi\phi} &=& \int d^2xd\tau \phi(\Psi^{\dagger}_{1i}\tau^{x}\Psi_{1 i}
+ \Psi^{\dagger}_{2i} \tau^{x}\Psi_{2i}).\label{Eq_S_lambda_rescaling}
\end{eqnarray}

The effective action represented by Eq.~(\ref{Eq_action_total}) was
studied recently in Ref.~\cite{Liu}. It was demonstrated that both
superfluid density and critical temperature $T_c$ are significantly
suppressed at nematic QCP $x_2$. However, the superconducting order
parameter $\psi$ was assumed in Ref.~\cite{Liu} to be classical, which
is valid only when $x_2$ is not close to $x_1$. In this paper, we go
beyond such approximation and consider the quantum fluctuations of
both superconducting and nematic order parameters.

\begin{figure}
   \includegraphics[width=4in]{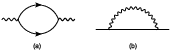}
   \centering
   \vspace{-0.3cm}
\caption{(a): Polarization function for nematic field $\phi$; (b):
Fermion self-energy correction due to nematic
fluctuation.}\label{Fig_Pi_Sigma}
\end{figure}

As emphasized in Ref.~\cite{Liu}, the gapless nodal qps can have
important impacts on the competition between superconductivity and
nematic order. The simplest way to include the fermionic degrees of
freedom is to introduce the polarization function $\Pi(q)$ due to
nodal qps into the effective action of nematic order, $S_{\phi}$. To
the leading order of $1/N$-expansion, the polarization function
$\Pi(q)$ is represented by the one-loop Feynman diagram shown in
Fig. \ref{Fig_Pi_Sigma}(a) and formally given by \cite{Huh, WLK}
\begin{eqnarray}
\Pi(\mathbf{q},\epsilon) = N\!\!\int\frac{d^{2}\mathbf{k}d\omega }{(2\pi)^3}
\mathrm{Tr}[\tau^{x}G_{0}(\mathbf{k},\omega)\tau^{x}
G_{0}(\mathbf{k+q},\omega+\epsilon)], \nonumber
\end{eqnarray}
where
\begin{eqnarray}
G_0(\mathbf{k},\omega) = \frac{1}{-i\omega + v_{F}k_x \tau^{z} +
v_{\Delta}k_y \tau^{x}} \nonumber
\end{eqnarray}
is the free propagator for nodal QPs $\Psi_1$ (free propagator for
nodal QPs $\Psi_2$ can be similarly written down). The polarization
function $\Pi(\epsilon,\mathbf{q})$ has already been calculated
previously \cite{Huh, Liu}, and is known to have the form
\begin{eqnarray}
\Pi(\mathbf{q},\epsilon) \!=\! \frac{N}{16v_F
v_\Delta}\!\!\left[\!\frac{\epsilon^2 \!+\! v_F^2 q_x^2}{\sqrt{\epsilon^2 \!+\! v_F^2
q_x^2 \!+\! v_{\Delta}^2 q_y^2}} \!+\! (q_x \!\leftrightarrow\!
q_y)\!\right]\!\!.
\end{eqnarray}
Including this term, the quadratic part of $S_{\phi}$ becomes
\begin{eqnarray}
\left(-2r+q^2\right)\phi^2 \rightarrow \left[-2r+ q^2 + \Pi(q)
\right]\phi^2.
\end{eqnarray}
From the expression of polarization $\Pi(q)$, it is easy to see that
inclusion of $\Pi(q)$ does not change the dynamical exponent $z = 1$
of $\phi$. However, $\Pi(q) \propto q$, so it dominates over the
kinetic term $q^2$ in the low energy regime. More importantly, the
polarization $\Pi(q)$ introduces two important quantities, nodal
qps' Fermi velocity $v_F$ and gap velocity $v_{\Delta}$, into the
effective action of $\phi$.

Although the polarization $\Pi(q)$ represents the influence of nodal
qps, we can not completely integrate the nodal qps out and drop them
from the effective theory. These gapless nodal qps should be
maintained for several reasons. First, according to the general
spirit of RG, one can safely integrate out high-energy modes at low
energies. However, the nodal qps are gapless and hence exist even at
the lowest energy. Integrating out gapless fermions completely may
lead to unphysical singularities \cite{Abanov, Rech, Metzner}.
Second, the coupling between gapless fermions and order parameter
fluctuation often causes non-Fermi liquid behavior in observable
quantities. In the case of nematic transition, the critical nematic
fluctuation gives rise to fermion velocity renormalization and
extreme anisotropy, which would be overlooked if fermions are fully
integrated out. As will be shown below, the velocity ratio $\kappa =
v_{\Delta}/v_{F}$ plays nontrivial roles.

The effective field theory contains seven parameters: $\alpha, r,
\beta, u, \gamma, v_F, v_\Delta$. They are all subjected to
renormalizations due to the mutual interactions among three field
operators: $\psi$, $\phi$, and $\Psi$. We will study the flow of
these seven parameters under scaling transformations and eventually
obtain seven RG equations. Since we study the $\psi$-$\phi$
interaction and $\Psi$-$\phi$ interaction on equal footings, these
seven RG equations are self-consistently coupled to each other. The
low-energy behaviors of these parameters and the possible fixed
points can be determined by solving these coupled RG equations.

\section{Renormalization group calculations}\label{sec_RG_analysis}

In this section, we make a RG analysis and obtain the flow equations
of all the aforementioned parameters. In order to examine the
impacts of gapless nodal qps, we go beyond the HMM theory and
maintain nodal qps throughout our calculations. We first analyze the
coupling between nematic order and nodal qps, and derive the RG
equations for fermion velocities, $v_{F,\Delta}$. We then consider
the competitive interaction between superconducting and nematic
order parameters, and obtain the RG equations for the rest five
parameters, $(\alpha,r,\beta,u,\gamma)$, which depend on the fermion
velocities, $v_{F,\Delta}$. These seven equations are
self-consistently coupled to each other since the fermion velocities
$v_{F,\Delta}$ appearing in the equations of
$(\alpha,r,\beta,u,\gamma)$ flow according to their own equations.

Our RG analysis will be performed within the framework presented in
Ref.~\cite{Shankar94}. According to RG theory \cite{Shankar94}, we
first employ the following scaling transformations:
\begin{eqnarray}
&&k_i=k'_ie^{-l},\\
&&\omega=\omega' e^{-l},\\
&&q_i=q'_ie^{-l},\\
&&\epsilon=\epsilon' e^{-l},
\end{eqnarray}
where $i=x,y$ and $l$ represents the scaling parameter. We next need
to determine how all the field operators flow under these
transformations. It is easy to know that the SC order parameter
$\psi$ should be re-scaled as
\begin{eqnarray}
\psi(\mathbf{k},\omega) = \psi'(\mathbf{k'},\omega')e^{5l/2}
\end{eqnarray}
on the basis of its action Eq. (\ref{Eq_S_psi}). Similarly, the rescaling
behavior of nodal qps $\Psi_{1,2}$ is found to be \cite{Huh}
\begin{eqnarray}
\Psi_{1,2}(\mathbf{k},\omega)
&=&\Psi'_{1,2}(\mathbf{k'},\omega')e^{(4 + C_1)l/2},
\end{eqnarray}
where the explicit expression of constant $C_1$ will be defined
below in Eq. (\ref{Eq_C1}).

However, determining the rescaling behavior of the nematic field
$\phi$ is much more complicated. In the standard RG theory
\cite{Shankar94}, one should regard the kinetic term of $\phi$ as
the fixed point and obtain the rescaling behavior of $\phi$ by
requiring such term invariant under scaling transformations.
Nevertheless, the kinetic term of nematic order $S_\phi$ is
irrelevant in the low-energy region, and therefore can not be
regarded as the fixed point. One has to choose another term to serve
as the fixed point.

As already demonstrated in Sec. \ref{sec_model}, the polarization
function $\Pi(q)$ generated by the strong interaction with gapless
nodal qps dominates over the kinetic term at low energies.
Naturally, one might expect to choose the polarization term, $\Pi(q)
\phi^2(q)$, to serve as the fixed point. However, we emphasize here
that it is not appropriate to obtain the scaling of $\phi$ by
requiring $\Pi(q)$ invariant under RG transformations. Eq.(9) tells
us that $\Pi(q)$ contains two fermion velocities, $v_{F, \Delta}$,
which are apparently scale-dependent and flow strongly as $l$ is
varying. If we insist on requiring the polarization term invariant
under scaling transformations, we have to assume that $v_{F,
\Delta}$ are always constants and do not flow with varying $l$.
Under this assumption, the important property of singular fermion
velocity renormalization cannot be properly taken into account in
our theoretical description of competing orders. For these reasons,
we believe it is not appropriate to adopt the polarization term as
the fixed point of the present model.

Since both the kinetic and polarization terms are not good choices
for the fixed point, we have to choose another term from the
effective action. It appears that the Yukawa coupling term, $\lambda
\phi \Psi^{\dagger}\tau^x \Psi$, is the only available candidate.
However, there is a very important problem: how to treat the
coupling constant $\lambda$? In the conventional perturbation
theory, usually one can make perturbative expansion in powers of
$\lambda$. Unfortunately, this scheme does not apply to the present
model because explicit RG calculations \cite{Vojta} have revealed
that $\lambda$ tends to diverge at the lowest energy. It was later
realized that a reasonable route to access such model is to fix
$\lambda$ at certain finite value \cite{Kim, Huh, Sachdevbook}, and
perform perturbative expansion in powers of $1/N$, where $N$ is
apparently the flavor. In this formalism, $\lambda$ is a constant
and does not flow with running $l$. One can first absorb $\lambda$
into $\phi$ \cite{Huh}, and then require the Yukawa coupling term
invariant under scaling transformations. It is now easy to know that
the nematic field transforms as \cite{Huh, Sachdevbook}
\begin{eqnarray}
\phi(\mathbf{q},\epsilon)=\phi'(\mathbf{q'},
\epsilon')e^{(2+C_3-C_1)l}, \label{Eq_phi_scaling}
\end{eqnarray}
where $C_{3}$ will be defined below in Eq. (\ref{Eq_C3}). This
expression is to be used in the following calculations.

\subsection{Flow equations of $v_{F}$ and $v_{\Delta}$}

The Yukawa-type interaction between nematic order and gapless nodal
qps has been recently investigated in several papers \cite{Kim, Huh,
Xu, Lawler2, Fritz, WLK}. It is well-known that the Fermi velocity
of nodal qps is indeed not equal to the gap velocity, i.e., $v_{F}
\neq v_{\Delta}$. Experiments \cite{Chiao, Orenstein00} have
determined that the velocity ratio $\kappa = v_{\Delta}/v_F \approx
0.1$. This ratio is a very important parameter since it enters into
a number of observable quantities of high-$T_c$ superconductors,
including electric conductivity \cite{Lee93, Durst}, thermal
conductivity \cite{Durst}, superfluid density \cite{Durst, Lee97},
and $T_c$ \cite{Lee97}. An interesting property revealed and
discussed in Refs.~\cite{Kim, Huh, Xu, Lawler2, Fritz, WLK} is that
the velocity anisotropy is significantly enhanced by the nematic
fluctuation.

The calculation of nodal qps self-energy function and the derivation
of flow equations have already been presented in previous
publications \cite{Huh, WLK}, and therefore are not shown here. It
is only necessary to summarize the basic calculations as well as the
relevant results. To the leading order, the fermion self-energy is
represented by the diagram Fig.~\ref{Fig_Pi_Sigma}(b), and has the
form
\begin{eqnarray}
\Sigma(\mathbf{k},\omega) = \int \frac{d^2\mathbf{q}d\epsilon}{(2\pi)^3}
G_0(\mathbf{k}+\mathbf{q},\omega+\epsilon)\frac{1}{\Pi(q)}.
\end{eqnarray}
As shown in Ref.~\cite{Huh}, it can be written as
\begin{eqnarray}
\frac{d\Sigma(\mathbf{k},\omega)}{d\ln\Lambda} =
C_1(-i\omega) + C_2 v_F k_x \tau^z + C_3 v_{\Delta} k_y \tau^x,
\end{eqnarray}
where
\begin{eqnarray}
C_1 &=& \frac{2(v_\Delta/v_F)}{N \pi^3}\int^{\infty}_{-\infty}dx
\int^{2\pi}_{0} d \theta \frac{x^2-\cos^2\theta-(v_\Delta/v_F)^2
\sin^2\theta}{\left[x^2+\cos^2\theta+(v_\Delta/v_F)^2
\sin^2\theta\right]^2}\mathcal {G}(x,\theta), \nonumber\label{Eq_C1} \\
C_2 &=& \frac{2(v_\Delta/v_F)}{N \pi^3}\int^{\infty}_{-\infty}dx
\int^{2\pi}_{0} d \theta  \frac{\cos^2\theta-x^2-(v_\Delta/v_F)^2
\sin^2\theta}{\left[x^2+\cos^2\theta+(v_\Delta/v_F)^2
\sin^2\theta\right]^2}
\mathcal{G}(x,\theta), \nonumber\label{Eq_C2}\\
C_3 &=& \frac{2(v_\Delta/v_F)}{N \pi^3}\int^{\infty}_{-\infty}dx
\int^{2\pi}_{0} d \theta \frac{x^2+\cos^2\theta-(v_\Delta/v_F)^2
\sin^2\theta}{\left[x^2+\cos^2\theta+(v_\Delta/v_F)^2
\sin^2\theta\right]^2}
\mathcal{G}(x,\theta), \nonumber\label{Eq_C3}\\
\mathcal{G}^{-1} &=& \frac{x^2+\cos^2\theta}{\sqrt
{x^2+\cos^2\theta+(v_\Delta/v_F)^2 \sin^2\theta}}
+ \frac{x^2+\sin^2\theta}{\sqrt{x^2 +
\sin^2\theta+(v_\Delta/v_F)^2\cos^2\theta}},\nonumber
\end{eqnarray}
Using Dyson equation, it is easy to get a renormalized fermion
propagator
\begin{eqnarray}
G^{-1}_{\psi}(\mathbf{k},\omega) = -i\omega + v_F k_x\tau^z +
v_{\Delta} k_{y}\tau^{x}  -\Sigma(\mathbf{k},\omega),
\end{eqnarray}
which leads to the following RG equations,
\begin{eqnarray}
\frac{d v_F}{dl}=(C_1 - C_2)v_F,\label{Eq_RG_v_F}\\
\frac{d v_\Delta}{dl}=(C_1 - C_3)v_\Delta,\label{Eq_RG_v_D}\\
\frac{d(v_{\Delta}/v_{F})}{dl} = (C_{2}-C_{3})(v_{\Delta}/v_{F}),\label{Eq_RG_v_D_v_F}
\end{eqnarray}
where $l>0$ is running scale. A straightforward analysis showed that
the ratio $v_{\Delta}/v_F$ flows to zero at the lowest energy,
giving rise to a novel fixed point of extreme velocity anisotropy
\cite{Huh}. Such fixed point in turn leads to a number of nontrivial
consequences, such as unusual broadening of spectral function
\cite{Kim}, non-Fermi liquid behavior \cite{Xu}, enhancement of dc
thermal conductivity \cite{Fritz}, and suppression of
superconductivity \cite{Liu}. To analyze the influence of velocity
renormalization and especially the extreme anisotropy manifested at
nematic QCP on the nature of superconducting transition, we require
that the constant fermion velocities appearing in the polarization
$\Pi(q)$ to flow with running scale $l$ according to Eqs.
(\ref{Eq_RG_v_F}) and (\ref{Eq_RG_v_D}).

\begin{figure}
   \includegraphics[width=4.5in]{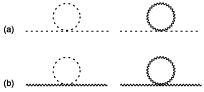}
   \centering
   \vspace{-0.3cm}
\caption{One-loop corrections to mass parameters $\alpha$ in (a) and
$r$ in (b), respectively.}\label{Fig_alpha_r}
\end{figure}

The extreme velocity anisotropy is a special feature of the nematic
QCP, where $r = 0$ and the nematic fluctuation is critical. Away
from nematic QCP, $r \neq 0$, so the nematic fluctuation leads only
to relatively unimportant renormalization of fermion velocities. For
$r \neq 0$, $v_{F,\Delta}$ and therefore their ratio
$v_{\Delta}/v_F$ remain finite.

\subsection{Flow equations of $\alpha$, $r$, $\beta$, $u$, and $\gamma$}

We next consider the competitive interaction between nematic and
superconducting order parameters. Our analysis follows closely the
scheme presented in a recent work of She \emph{et al.}
\cite{Zaanen}. It was assumed in Ref.~\cite{Zaanen} that all the
fermionic degrees of freedom can be integrated out and their effects
can be represented by the dynamical exponent $z$. Compared with
Ref.~\cite{Zaanen}, the main difference here is the inclusion of
polarization $\Pi(q)$ in the effective action of nematic order
$\phi$, which is supposed to reflect the influence of nodal qps. The
corresponding (sub)action that describes ordering competition is
\begin{eqnarray}
S_{\mathrm{com}}=S_{\psi}+S_{\phi}+S_{\psi\phi},\label{Eq_action_eff}
\end{eqnarray}
where
\begin{eqnarray}
S_\psi &=& \frac{1}{2}\int \frac{d^2\mathbf{k}d\omega}{(2\pi)^3}
\left(-2\alpha+\mathbf{k}^2+\omega^2\right)\psi^2\nonumber\\
&& + \frac{\beta}{2}\int
\prod^4_{m=1}\frac{d^2\mathbf{k}_md\omega_m}
{(2\pi)^3}\delta^2\left(\sum\mathbf{k}_m\right)
\delta\left(\sum\omega_m\right)\psi^4,\nonumber\\
S_\phi &=& \int \frac{d^2\mathbf{q}d\epsilon}{(2\pi)^3}
\frac{1}{2}\left[-2r+\mathbf{q}^2+\epsilon^2+\Pi(q)\right]\phi^2\nonumber\\
&&+\frac{u}{2} \int \prod^4_{m=1}\frac{d^2\mathbf{q}_md\epsilon_m}
{(2\pi)^3}\delta^2\left(\sum\mathbf{q}_m\right)
\delta\left(\sum\epsilon_m\right)\phi^4,\nonumber\\
S_{\psi\phi} &=& \gamma\int \prod_{i=1,2}
\frac{d^2\mathbf{k}_id\omega_id^2\mathbf{q}_id\epsilon_i}
{(2\pi)^6}\psi(\mathbf{k}_i,\omega_i)
\phi(\mathbf{q}_i,\epsilon_i)\nonumber\\
&&\times \delta^2\left(\mathbf{k}_1+\mathbf{k}_2+ \mathbf{q}_1 +
\mathbf{q}_2\right) \delta\left(\omega_1+\omega_2+\epsilon_1
+\epsilon_2\right).\nonumber
\end{eqnarray}
Before performing a standard RG analysis within this action, it is
convenient to rescale momenta and energy by $\Lambda$, i.e,
$\mathbf{k}\rightarrow \mathbf{k}/\Lambda$, $\omega \rightarrow
\omega/\Lambda$.

Each field operator can be separated into slow mode and fast mode,
i.e.,
\begin{eqnarray}
\psi &=& \psi_s + \psi_f, \\
\phi &=& \phi_s + \phi_f.
\end{eqnarray}
After introducing an UV cutoff $\Lambda$, we can define the slow
mode of superconducting order parameter as $\psi_s = \psi(k)$ with
$0 < k < e^{-l}\Lambda$ and the fast mode as $\psi_f = \psi(k)$ with
$e^{-l}\Lambda < k < \Lambda$, using the formalism of
Ref.~\cite{Shankar94}. Based on such modes separation, the effective
action (\ref{Eq_action_eff}) is decomposed into three parts: $S^{s}$
that contains only slow modes, $S^{f}$ that contains only fast
modes, and $S^{sf}$ that contains both slow and fast modes. More
concretely, we have
\begin{eqnarray}
S_{\mathrm{com}}
&=&S^{s}+S^{f}+S^{sf}\nonumber\\
&=&\left(S^{s}_\psi+S^{s}_\phi+S^{s}_{\psi\phi}\right)
+\left(S^{f}_\psi+S^{f}_\phi\right) +
S^{sf},\label{Eq_action_eff_s_f}
\end{eqnarray}
where
\begin{eqnarray}
S^{s}_{\psi} &=& \frac{1}{2}\int
\frac{d^2\mathbf{k}d\omega}{(2\pi)^3} \left(-2\alpha+\mathbf{k}^2 +
\omega^2\right)\psi_s^2 \nonumber\\
&& + \frac{\beta}{2}\int \prod^4_{m=1}
\frac{d^2\mathbf{k}_md\omega_m}{(2\pi)^3}
\delta^2\left(\sum\mathbf{k}_m\right)
\delta\left(\sum\omega_m\right)\psi_s^4,\nonumber\\
S^{s}_{\phi} &=& \frac{1}{2}\int
\frac{d^2\mathbf{q}d\epsilon}{(2\pi)^3}
\left[-2r+\mathbf{q}^2+\epsilon^2+\Pi(q)\right]\phi_s^2 \nonumber\\
&& + \frac{u}{2}\int \prod^4_{m=1}\frac{d^2\mathbf{q}_md\epsilon_m}
{(2\pi)^3}\delta^2\left(\sum\mathbf{q}_m\right)
\delta\left(\sum\epsilon_m\right)\phi_s^4,\nonumber\\
S^{s}_{\psi\phi} &=& \gamma\int \prod_{i=1,2}
\frac{d^2\mathbf{k}_id\omega_id^2\mathbf{q}_id\epsilon_i}
{(2\pi)^6}\psi(\mathbf{k}_i,\omega_i)
\phi(\mathbf{q}_i,\epsilon_i)\nonumber\\
&& \times \delta^2 \left(\mathbf{k}_1+\mathbf{k}_2+
\mathbf{q}_1+\mathbf{q}_2\right)\delta\left(\omega_1+\omega_2+\epsilon_1
+\epsilon_2\right);\nonumber\\
S^{f}_{\psi} &=& \frac{1}{2}\int
\frac{d^2\mathbf{k}d\omega}{(2\pi)^3}
\left(-2\alpha+\mathbf{k}^2+\omega^2\right)\psi_f^2\nonumber\\
&&+\frac{\beta}{2}\int \prod^4_{m=1}\frac{d^2\mathbf{k}_md\omega_m}
{(2\pi)^3}\delta^2\left(\sum\mathbf{k}_m\right)
\delta\left(\sum\omega_m\right)\psi_f^4,\nonumber\\
S^{f}_{\phi} &=& \frac{1}{2} \int
\frac{d^2\mathbf{q}d\epsilon}{(2\pi)^3}
\left[-2r+\mathbf{q}^2+\epsilon^2+\Pi(q)\right]\phi_f^2\nonumber\\
&&+\frac{u}{2}\int \prod^4_{m=1}\frac{d^2\mathbf{q}_md\epsilon_m}
{(2\pi)^3}\delta^2\left(\sum\mathbf{q}_m\right)
\delta\left(\sum\epsilon_m\right)\phi_f^4,\nonumber\\
S^{sf} &=& \int \frac{d^2\mathbf{k}d\omega}{(2\pi)^3}
\left[(3\beta)\psi_f \psi_f \psi_s \psi_s + (3u)\phi_f
\phi_f \phi_s \phi_s\right] \nonumber\\
&& + \int \prod_{i=1,2}\frac{d^2\mathbf{k}_i d\omega_i
d^2\mathbf{q}_id\epsilon_i}{(2\pi)^6} \delta^2
\left(\mathbf{k}_1+\mathbf{k}_2+\mathbf{q}_1+\mathbf{q}_2\right)\nonumber\\
&& \times \delta\left(\omega_1+\omega_2+\epsilon_1+\epsilon_2\right)
\left(\gamma\psi_s\psi_s\phi_f\phi_f\right.\nonumber\\
&&\left.+\gamma\psi_f\psi_f\phi_s\phi_s+ 4\gamma\psi_f\psi_s\phi_f\phi_s\right).\nonumber
\end{eqnarray}
After this decomposition, the partition function can be rearranged
in the following way,
\begin{eqnarray}
Z&=&\int \mathcal{D}\psi_s\mathcal{D}\phi_s \mathcal{D}\bar{\Psi} \mathcal{D}\Psi
\int \mathcal{D}\psi_f\mathcal{D}\phi_f\nonumber\\
&&\times\exp\left[{\left(S^{s}_\psi+S^{s}_\phi+S^{s}_{\psi\phi}\right)
+\left(S^{f}_\psi+S^{f}_\phi\right) +S^{sf}}\right]\nonumber\\
&=&\int \mathcal{D}\psi_s\mathcal{D}\phi_s \mathcal{D}\bar{\Psi} \mathcal{D}\Psi
\exp{\left(S^{s}_\psi+S^{s}_\phi+S^{s}_{\psi\phi}\right)}\nonumber\\
&&\times \int \mathcal{D}\psi_f\mathcal{D}\phi_f
\exp\left[{\left(S^{f}_\psi+S^{f}_\phi\right)
+S^{sf}}\right] \nonumber\\
&=&\int \mathcal{D}\psi_s\mathcal{D}\phi_s \mathcal{D}\bar{\Psi}
\mathcal{D}\Psi \exp{\left(S'^{s}_\psi+S'^{s}_\phi +
S'^{s}_{\psi\phi}\right)}\nonumber\\
&\equiv&\int \mathcal{D}\psi_s\mathcal{D}\phi_s \mathcal{D}\bar{\Psi}
\mathcal{D}\Psi\exp{\left(S'_{\mathrm{eff}}\right)}.
\end{eqnarray}
The next step is to integrate over all the fast modes, and obtain an
effective action of slow modes. The functional integration can be
performed using the standard diagrammatic techniques. The
propagators for the superconducting order $\psi$ and the nematic order $\phi$,
shown in Fig.~\ref{propagators}, are
\begin{eqnarray}
G_\psi(\mathbf{k},\omega) &=&
\frac{1}{-2\alpha+\mathbf{k}^2+\omega^2},\\
G_\phi(\mathbf{q},\epsilon) &=&
\frac{1}{-2r+\mathbf{q}^2+\epsilon^2+\Pi(q)}.
\end{eqnarray}
The polarization appearing in $G_\phi(\mathbf{q},\epsilon)$ reflects
the presence of gapless nodal qps. As already pointed out, $\Pi(q)$
dominates over the kinetic term $q^2$ in the low-energy regime. In
order to further simplify the nematic propagator, we consider the
close vicinity of nematic QCP where $r$ is very small. In this case,
we are allowed to approximate the nematic propagator by
\begin{eqnarray}
G_\phi(\mathbf{q},\epsilon)
&=& \frac{1}{-2r +\mathbf{q}^2+\epsilon^2+ \Pi(q)} \nonumber \\
&\approx& \frac{1}{-2r + \Pi(q)}\hspace{0.5cm}
\mathrm{(at \,\, low\,\, energy)} \nonumber \\
&\approx& \frac{1}{\Pi(q)\left[1 - \frac{2r}{\Pi(q)}\right]} \nonumber \\
&\approx&\frac{1}{\Pi(q)}+\frac{2r}{\Pi^2(q)}.
\end{eqnarray}

\begin{figure}
   \includegraphics[width=4.25in]{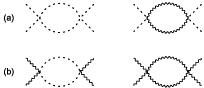}
   \centering
   \vspace{-0.10cm}
\caption{One-loop corrections to quartic coefficients $\beta$ in (a)
and $u$ in (b), respectively.}\label{Fig_beta_u}
\end{figure}

Now let us work in the spherical coordinates. We first define
$\epsilon \equiv vq_0 = vq\cos\theta$, $q_1 =
q\sin\theta\cos\varphi$, and $q_2 = q\sin\theta\sin\varphi$, with
$v$ being the velocity of nematic order parameter $\phi$. For
convenience, we could extract constant $v$ from the energy $\omega$
of superconducting order parameter $\psi$. Accordingly, the velocities
of nodal qps, $v_F$ and $v_\Delta$, can be divided by $v$. After this
manipulation, $v_F$ and $v_\Delta$ become dimensionless. Now the
polarization function can be written in the form,
\begin{eqnarray}
\Pi(q,\theta,\varphi) = q D(\theta,\varphi),\label{Eq_Pi_D}
\end{eqnarray}
where the function
\begin{eqnarray}\label{Eq_D_theta_varphi}
D(\theta,\varphi) &=& \frac{N}{16v_F v_{\Delta}}
\left(\frac{\cos^2\theta + v_F^2 \sin^2\theta\cos^2\varphi}
{\sqrt{\cos^2\theta + v_F^2 \sin^2\theta\cos^2\varphi + v_{\Delta}^2
\sin^2\theta\sin^2\varphi}}\right. \nonumber\\
&& +\left.\frac{\cos^2\theta + v_F^2\sin^2\theta\sin^2\varphi}
{\sqrt{\cos^2\theta + v_F^2\sin^2\theta\sin^2\varphi +
v_{\Delta}^2\sin^2\theta\cos^2\varphi}}\right)\nonumber
\end{eqnarray}
is dimensionless. Before proceeding with the next calculations, it
is helpful to define
\begin{eqnarray}
F_1 &\equiv& F_1(v_F,v_\Delta)=
\int^\pi_0\!\!d\theta\int^{2\pi}_0\!\!d\varphi
\frac{\sin\theta}{(2\pi)^{3} D(\theta,\varphi)},\label{Eq_F_1} \\
F_2 &\equiv& F_2(v_F,v_\Delta)=
\int^\pi_0\!\!d\theta\int^{2\pi}_0\!\!d\varphi
\frac{\sin\theta}{(2\pi)^{3} D^2(\theta,\varphi)},\label{Eq_F_2}\\
F_3 &\equiv& F_3(v_F,v_\Delta)=
\int^\pi_0\!\!d\theta\int^{2\pi}_0\!\!d\varphi
\frac{\sin\theta}{(2\pi)^{3} D^3(\theta,\varphi)}.\label{Eq_F_3}
\end{eqnarray}
With these arrangements, we can now turn to calculate the one-loop
contribution to the RG equations of all parameters.

\subsubsection{ $\alpha,r$}

The diagrams contributing to $\alpha$ to leading order are shown in
Fig. \ref{Fig_alpha_r}(a). We perform the following calculations,
\begin{eqnarray}
S[\alpha]
&=&\int^{b}\frac{d^{3}q}{(2\pi)^{3}}
\psi_s(q)\psi_s(-q)\left[-2\alpha+3\beta\int^1_{b}
\frac{d^{3}q'}{(2\pi)^{3}} G^f_\psi+\gamma\int^1_b\frac{d^{3}q'}{(2\pi)^{3}}
G^f_\phi\right]\nonumber\\
&\approx&\int^{b}\frac{d^{3}q}{(2\pi)^{3}}
\psi_s(q)\psi_s(-q)(-2\alpha)\nonumber\\
&&\times\exp\left\{-\frac{l}{2\alpha}\left[\frac{3\beta}{2\pi^2}
\left(1+2\alpha\right) + \gamma\left(F_1 + 2rF_2\right)\right]\right\}\nonumber\\
&=&\int^{1}\frac{d^{3}q}{(2\pi)^{3}}e^{-3l}
\psi_s(q)\psi_s(-q)e^{5l/2}e^{5l/2}(-2\alpha)\nonumber\\
&&\times\exp\left\{-\frac{l}{2\alpha}\left[\frac{3\beta}{2\pi^2}
\left(1+2\alpha\right) +
\gamma(F_1+2rF_2)\right]\right\}\nonumber\\
&\equiv&\int^{1}\frac{d^{3}q}{(2\pi)^{3}}
\psi_s(q)\psi_s(-q)
(-2\alpha'),\label{Eq_S_alpha}
\end{eqnarray}
where
\begin{eqnarray}
\alpha' = \alpha
\exp\left\{2l-\frac{l}{2\alpha}\left[\frac{3\beta}{2\pi^2}\left(
1+2\alpha\right)+\gamma (F_1+2rF_2)\right]\right\}.\nonumber
\end{eqnarray}
Its derivative with respect to running scale $l$ is
\begin{eqnarray}
\frac{d\alpha}{dl} = 2\alpha-\left[\frac{3\beta}{4\pi^2}
\left( 1+2\alpha\right) + \frac{\gamma}{2} (F_1+2rF_2)\right].
\end{eqnarray}

By calculating the diagrams shown in Fig. \ref{Fig_alpha_r}(b), the
flow equation for parameter $r$ can be obtained similarly,
\begin{eqnarray}
\frac{dr}{dl}&=&[1+2(C_3-C_1)]r-\left[\frac{3u}{2}(F_1+2rF_2)
+\frac{\gamma}{4\pi^2}\left( 1+2\alpha\right)\right].
\end{eqnarray}

In the absence of fermionic degrees of freedom, $F_1 = F_2 =
\frac{1}{2\pi^2}$ and $[1+2(C_3-C_1)]r$ replaced with $2r$, then the
flow equation of $\alpha$ is identical to that of $r$ \cite{Zaanen}.
Such an ``exchange symmetry" is certainly broken by gapless nodal
qps via the polarization $\Pi(q)$ appearing in the effective action
of nematic order $\phi$.

\subsubsection{ $\beta,u$}

The one-loop corrections to $\beta$ are depicted in
Fig.~\ref{Fig_beta_u}(a). By paralleling the steps performed
in Eq. (\ref{Eq_S_alpha}), we can similarly obtain
\begin{eqnarray}
S[\beta]&\equiv&\int^{1} \prod^4_{m=1}
\frac{d^2\mathbf{k}_md\omega_m}
{(2\pi)^3}\delta^2\left(\sum\mathbf{k}_m\right)
\delta\left(\sum\omega_m\right)
\frac{\beta'}{2}|\psi_s|^4,\nonumber
\end{eqnarray}
where
\begin{eqnarray}
\beta'=\beta \exp{\left\{1-\frac{2}{\beta}
\left[\frac{9\beta^2}{2\pi^2}
+\gamma^2(F_2+4rF_3)\right]\right\}l}.\nonumber
\end{eqnarray}
It leads to
\begin{eqnarray}
\frac{d\beta}{dl} = \beta- \left[\frac{9\beta^2}{\pi^2} +
2\gamma^2(F_2 + 4rF_3)\right].
\end{eqnarray}

By calculating diagrams shown in Fig.~\ref{Fig_beta_u}(b), the RG
equation for $u$ is found to be
\begin{eqnarray}
\frac{du}{dl} &=& [-1+4(C_3-C_1)]u  -\left[\frac{\gamma^2}{\pi^2}
+ 18u^2(F_2+4rF_3)\right].
\end{eqnarray}
Once again, the influence of gapless nodal qps is reflected in the
functions $F_{2,3}$ and $C_{1,3}$.

\subsubsection{ $\gamma$}

\begin{figure}
   \includegraphics[width=5in]{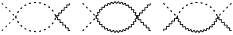}
   \centering
   \vspace{-0.10cm}
\caption{One-loop corrections to coupling constant
$\gamma$.}\label{Fig_gamma}
\end{figure}

Now we consider the flow of $\gamma$, which characterizes the
strength of competitive interaction. The one-loop corrections to the
term $\gamma \psi^2 \phi^2$ has three diagrams, presented in
Fig.~\ref{Fig_gamma}. Following similar procedure, we eventually
obtain
\begin{eqnarray}
S[\gamma]\!&\equiv&\!\!\int^{1}\!\!
\prod^4_{m=1}\frac{d^2\mathbf{k}_m}
{(2\pi)^2}\frac{d\omega_m}{2\pi}\delta^2
\left(\sum\mathbf{k}_m\right)
\delta\left(\sum\omega_m\right)
\gamma'\psi^2_s\phi^2_s,\nonumber
\end{eqnarray}
where
\begin{eqnarray}
\gamma'&=&\gamma\exp\Bigl\{l-\Bigl[\frac{3\beta}{\pi^2}+6u(F_2+4rF_3)
+8\gamma\left(F_1+2rF_2+2\alpha F_1\right)\Bigr]l\Bigr\}.\nonumber
\end{eqnarray}

The flow equation of $\gamma$ is therefore given by
\begin{eqnarray}
\frac{d \gamma}{dl}&=&\gamma\Bigl\{2(C_3-C_1)
-\Bigl[\frac{3\beta}{\pi^2}+6u(F_2+4rF_3)\nonumber\\
&&+8\gamma\left(F_1+2rF_2+2\alpha F_1\right)\Bigr]\Bigr\}.
\end{eqnarray}

\section{Numerical results and physical implications}\label{sec_numerical_discussion}

In the last section, we have already obtained the RG equations for a
number of relevant parameters. In order to specify the possible
fixed point of the system under consideration, we need to solve
these RG equations.

For later reference, it is useful to list all the RG equations
obtained in the last section,
\begin{eqnarray}
\frac{d\alpha}{dl} &=&2\alpha-\left[\frac{3\beta}
{4\pi^2}\left(1+2\alpha\right)
+\frac{\gamma}{2}(F_1+2rF_2)\right],\label{Eq_alpha_RGEq}\\
\frac{dr}{dl}&=&[1+2(C_3-C_1)]r-\left[\frac{3u}{2}(F_1+2rF_2)+\frac{\gamma}{4\pi^2}
\left( 1+2\alpha\right)\right],\label{Eq_r_RGEq}\\
\frac{d\beta}{dl}&=&\beta-\left[\frac{9\beta^2}{\pi^2}
+2\gamma^2(F_2+4rF_3)\right],\label{Eq_beta_RGEq}\\
\frac{du}{dl}&=&[-1+4(C_3-C_1)]u-\left[\frac{\gamma^2}{\pi^2}
+18u^2(F_2+4rF_3)\right],\label{Eq_u_RGEq}\\
\frac{d\gamma}{dl}&=&\gamma\Bigl\{2(C_3-C_1)
-\Bigl[\frac{3\beta}{\pi^2}+6u(F_2+4rF_3)\nonumber\\
&&+8\gamma\left(F_1+2rF_2+2\alpha
F_1\right)\Bigr]\Bigr\},\label{Eq_gamma_RGEq}\\
\frac{d v_F}{dl} &=& (C_1 - C_2)v_F,\label{Eq_RG_vF}\\
\frac{d v_\Delta}{dl} &=& (C_1 - C_3)v_{\Delta}.\label{Eq_RG_vD}
\end{eqnarray}

Compared with the case in which the action contains only two bosonic
order parameters, the gapless nodal qps show their existence by
entering into the three functions $F_{1,2,3}$, which are all
functions of fermion velocities, $v_{F,\Delta}$. We now address the
influence of these nodal qps on the fixed-point properties of the
interacting system.

For finite $r$, the quantum fluctuation of nematic order and its
coupling to gapless nodal qps are relatively weak, and hence only
lead to unimportant renormalizations of fermion velocities.
Actually, the nematic fluctuation is singular only in the close
vicinity of the nematic QCP where $r = 0$. Here, we focus on the
nematic QCP, and solve the above RG equations self-consistently
after taking into account the singular velocity renormalization of
nodal qps driven by the critical nematic fluctuation.

\subsection{Theoretical analysis}

Before solving the RG equations, it is useful to first make a simple
theoretical analysis of the possible scaling behavior. At the
nematic QCP, $r = 0$, so the RG equations can be simplified to
\begin{eqnarray}
\frac{d\alpha}{dl} &=& 2\alpha\!-\!\left[\frac{3\beta}{4\pi^2}
\left(
1+2\alpha\right) \!+\! \frac{\gamma }{2}F_1\right],\\
\frac{d\beta}{dl} &=& \beta\!-\!\left[\frac{9\beta^2}{\pi^2}
\!+\!2\gamma^2F_2\right],\\
\frac{du}{dl} &=& [-1+4(C_3-C_1)]u\!-\!\left[\frac{\gamma^2}{\pi^2}
\!+\!18u^2F_2\right],\\
\frac{d\gamma}{dl} &=& \gamma\left\{\!2(C_3\!-\!C_1)\!
-\!\left[\!\frac{3\beta}{\pi^2}\!+\!6uF_2\!+\!8\gamma F_1\left(1 +
2\alpha \right)\!\right]\!\right\}. \label{Eq_RG_gamma1}
\end{eqnarray}
Here, we are particularly interested in the behavior of the
competitive coupling constant $\gamma$. The initial values of two
quartic coefficients, $\beta$ and $u$, are taken to be positive to
ensure the system stable. Numerical computations show that
$F_{1,2,3}$ are all positive and is driven to vanish as $l
\rightarrow \infty$ due to the extreme velocity anisotropy, namely
$v_{\Delta}/v_F \rightarrow 0$, at the nematic QCP. Moreover, since
$C_3 - C_1 \rightarrow 0$ as $l \rightarrow \infty$, the right hand
side of Eq.~(\ref{Eq_RG_gamma1}) is actually negative in the
low-energy region. As a result, the parameter $\gamma$ is expected
to be strongly irrelevant at low energy. It thus turns out that the
superconducting and nematic orders might be decoupled, which is
directly owing to the presence of gapless nodal qps.

To examine the effects of gapless nodal qps, here we present a set
of RG equations
\begin{eqnarray}
\frac{d\alpha}{dl}&=&2\alpha-\frac{1}{8\pi^2}\left[3\beta
\left( 1+2\alpha\right)+\gamma\right],\\
\frac{d\beta}{dl}&=&\beta-\frac{1}{4\pi^2}
\left(9\beta^2+\gamma^2\right),\\
\frac{du}{dl}&=&u-\frac{1}{4\pi^2}\left(9u^2+\gamma^2\right),\\
\frac{d\gamma}{dl}&=&\gamma\left[1-\frac{1}{4\pi^2}
\left(3\beta+3u+4\gamma\right)\right],\label{Eq_RG_gamma2}
\end{eqnarray}
which are obtained without considering fermionic degrees of freedom
\cite{Zaanen}. It is easy to see that the right hand side of Eq.
(\ref{Eq_RG_gamma2}) is not always negative. These equations have
been analyzed in Ref.~\cite{Zaanen}. It was found that the fixed
point structure depends on the initial values of $\beta$ and $u$.
For certain values of $\beta$ and $u$, the system has a stable fixed
point with coupling parameter $\gamma$ approaches a finite constant
$\gamma*$. In such case, the superconducting and nematic orders
experience a strong competitive interaction, and tend to suppress
each other. For other values of $\beta$ and $u$, the system has no
stable fixed point, which implies the instability of the system and
the appearance of first order transition. In any case, the
properties of the system without including nodal qps are
significantly different from that obtained after including nodal
qps.

The above analysis strongly suggests that gapless nodal qps can have
a significant influence on the fixed points of the present
interacting system. In particular, including the dynamics of nodal
qps may fundamentally change the nature of the competitive
interaction between the superconducting and nematic orders.

\subsection{Numerical results and physical implications}

In order to confirm our qualitative analysis, we now would solve
the RG equations numerically. As already pointed out, the fermion
velocities $v_{F,\Delta}$ are no longer constants at the nematic
QCP: they become scale-dependent and their ratio $v_{\Delta}/v_F$
vanishes at the lowest energy. In this case, the fixed point should
be obtained by self-consistently solving all the RG equations
presented in Eqs. (\ref{Eq_alpha_RGEq}, \ref{Eq_r_RGEq}, \ref{Eq_beta_RGEq},
\ref{Eq_u_RGEq}, \ref{Eq_gamma_RGEq}, \ref{Eq_RG_vF}, \ref{Eq_RG_vD}).
However, since the equations of $v_{F,\Delta}$ are relatively independent
of all the others, they can be solved first. It is easy to get $v_F^*=0,
v_\Delta^*=0$, in the limit $l \rightarrow \infty$. This immediately
implies that $F_1 = F_2 = F_3 = 0$. Now the rest set of equations
become much simpler, and given by
\begin{eqnarray}
\frac{d\alpha}{dl}&=&2\alpha-\frac{3\beta}{4\pi^2}
\left( 1+2\alpha\right)=0,\\
\frac{d\beta}{dl} &=& \beta-\frac{9\beta^2}{\pi^2}=0,\\
\frac{du}{dl} &=& -u-\frac{\gamma^2}{\pi^2}=0,\\
\frac{d\gamma}{dl} &=& -\gamma\frac{3\beta}{\pi^2} = 0.
\end{eqnarray}

These coupled equations have two solutions:
\begin{eqnarray}
1)\qquad r^*=u^*=\gamma^* = 0, \qquad \beta^*=\alpha^*=0; \\
2)\qquad r^*=u^*=\gamma^*=0, \qquad \beta^*=\frac{\pi^2}{9}, \qquad
\alpha^*=\frac{1}{22}.
\end{eqnarray}
One can check that the second solution corresponds to a stable fixed
point, and that the first one is unstable. It is important to notice
that $\gamma^* = 0$ at the stable fixed point, which means the
competitive interaction between the superconducting and nematic
order parameters are actually irrelevant. These results indicate that
these two long-range orders are decoupled from each other at the
nematic QCP. Hence, we can draw two conclusions. First, due to this
decoupling, the superconducting and nematic long-range orders can
coexist homogeneously. Second, both the superconducting and nematic
phase transitions remain continuous, which is apparently different
from the results of first order transitions obtained without
considering fermionic degrees of freedom.

It is now interesting to further discuss the role played by gapless
nodal qps. Currently, the quantum critical phenomena are usually
investigated within HMM theory, which supposes that all the
fermionic degrees of freedom can be entirely integrated out. If we
use this scheme in our case, we would obtain an effective action
that consists of solely two bosonic order parameters, $\psi$ and
$\phi$. The gapless nodal qps only show their existence in the
polarization $\Pi(q)$, which contributes a $\propto \Pi(q)\phi^2$
term to the effective Lagrangian of $\phi$. The fermion velocities,
$v_{F,\Delta}$, have to take bare values and can not be
renormalized, because the coupling between nematic order and nodal
qps can not be properly accounted for once the nodal qps are
completely integrated out. It would not be possible to incorporate
the extreme velocity anisotropy driven by critical nematic
fluctuation into the theoretical analysis. However, as demonstrated
in the above calculations, such extreme velocity anisotropy does
have significant effects on the RG trajectories. It is therefore
necessary to include the dynamics of nodal qps.

In the absence of nodal qps, the propagator of nematic order
parameter behaves as $G_{\phi} \propto 1/q^2$. This propagator is
strongly singular in the small momenta limit $q \rightarrow 0$.
After including gapless nodal qps, the nematic propagator becomes
$G_{\phi} \propto 1/q$, which is less singular compared with
$G_{\phi} \propto 1/q^2$ as $q \rightarrow 0$. It is clear that the
coupling with nodal qps weakens the critical fluctuation of nematic
order parameter, which in turn leads to the decoupling between the
nematic and superconducting orders. Apparently, gapless nodal qps do
play an important role and thus should be seriously considered in
the effective theory of competing orders.

\section{Summary and discussion}\label{sec_summary}

In summary, we have carried out a RG analysis within an effective
low-energy field theory that describes the interplay between
superconductivity and nematic order in the context of $d$-wave
high-$T_c$ superconductors. Different from some previous theoretical
treatments, we go beyond the HMM framework and incorporate gapless
nodal qps explicitly in our calculations. After analyzing the RG
equations of a number of physical parameters, we have demonstrated
that the gapless nodal qps have significant impacts on the interplay
between \emph{d}-wave superconductivity and nematic order. If the
nodal qps are entirely neglected, the competition between
superconducting and nematic orders can result in runaway behavior,
which in turn drives first order transition \cite{Zaanen}. However,
including the dynamics of nodal qps can change this picture
fundamentally, and give rise to a stable fixed point with these two
bosonic order parameters decoupled from each other in the vicinity
of nematic QCP. Therefore, both the superconducting and nematic
phase transitions remain continuous. Moreover, the \emph{d}-wave
superconductivity can coexist with the nematic order homogeneously.
These results indicate that it should be important to include the
dynamics of gapless fermions in the theoretic description of
competing orders.

Competition between superconductivity and nematic order is just a
simple example of the rich phenomena of ordering competition in
unconventional superconductors. It would be more interesting to
investigate the competition between superconductivity and
antiferromagnetism, which is believed to be a fundamental issue not
only in high-$T_c$ cuprate superconductors but also in heavy fermion
and iron-based superconductors. Compared with the case of competing
nematic order, the interplay between superconductivity and
antiferromagnetism is more complicated and is expected to exhibit
more interesting behaviors. For instance, the antiferromagnetic
order parameter is a complex scalar field and carries a finite wave
vector $\mathbf{Q}$ which is often incommensurate. Furthermore, the
antiferromagnetic order parameter may acquire a nontrivial dynamical
exponent, $z \neq 1$, due to its coupling to gapless fermions, which
would make RG calculations more involved \cite{Zaanen}.
Nevertheless, despite the technical difficulties, the general
formalism presented in this paper can be applied to analyze the
effects of fermionic degrees of freedom on the interplay between
superconductivity and antiferromagnetism.

\ack{We are grateful to Jian-Huang She for very helpful
communications, and to Jing-Rong Wang for recent collaboration on
relevant projects. G.Z.L. acknowledges support by the National
Natural Science Foundation of China under grants No. 11074234 and
No. 11274286, and the Visitors Program of MPIPKS at Dresden.}


\section*{References}

\begin{thebibliography}{99}


\bibitem{Kivelson03}
Kivelson S A, Bindloss I P, Fradkin E, Oganesyan V, Tranquada J M,
Kapitulnik A and Howald C 2003 \emph{Rev. Mod. Phys.} {\bf 75},
1201

\bibitem{Vojta}
Vojta M 2009 \emph{Adv. Phys.} {\bf 58}, 699

\bibitem{Fradkin}
Fradkin E, Kivelson S A, Lawler M J, Eisenstein J P and Mackenzie A P,
2010 \emph{Annu. Rev. Condens. Matter Phys.} {\bf 1}, 153;
Fradkin E 2012 in \emph{Proceedings of the Les Houches Summer School on "Modern theories
of correlated electron systems", Les Houches, Haute Savoie, France (May 2009)},
Lecture Notes in Physics 843, editored by D. C. Cabra, A. Honecker, and P. Pujol,
(Springer-Verlag, Berlin).

\bibitem{heavyfermion}
Gegenwart P, Si Q and Steglich F 2008 \emph{Nat. Phys.} {\bf 4}, 186;
Stockert O, Kirchner S, Steglich F  and Si Q 2012 \emph{J. Phys.
Soc. Jpn.} {\bf 81}, 011001

\bibitem{Flouquet}
Knebel G, Aoki D and Flouquet J 2009 \emph{arXiv}:0911.5223.

\bibitem{Arovas}
Arovas D P, Berlinsky A J, Kallin C and Zhang S -C
1997 \emph{Phys. Rev. Lett.} {\bf 79}, 2871

\bibitem{Demler}
Demler E, Sachdev S, and Zhang Y 2001 \emph{Phys. Rev. Lett.} {\bf 87},
067202

\bibitem{Kivelson02}
Kivelson S A, Lee D -H, Fradkin E and Oganesyan V 2002
\emph{Phys. Rev. B} {\bf 66}, 144516

\bibitem{Chubukov}
Vorontsov A B, Vavilov M G and Chubukov A V 2009 \emph{Phys. Rev. B}
{\bf 79}, 060508(R)

\bibitem{Nussinov}
Nussinov Z, Vekhter I and Balatsky A V 2009 \emph{Phys. Rev. B} {\bf 79},
165122

\bibitem{Millis10}
Millis A J 2010 \emph{Phys. Rev. B} {\bf 81}, 035117

\bibitem{Schmalian}
Fernandes R M  and Schmalian J 2010 \emph{Phys. Rev. B} {\bf 82}, 014521

\bibitem{Moon}
Moon E G and Sachdev S 2010 \emph{Phys. Rev. B} {\bf 82}, 104516

\bibitem{Zaanen}
She J -H, Zaanen J, Bishop A R and Balatsky A V 2010 \emph{Phys. Rev. B}
{\bf 82}, 165128

\bibitem{Chowdhury}
Chowdhury D, Berg E and Sachdev S 2011 \emph{Phys. Rev. B} {\bf 84}, 205113

\bibitem{Lake1}
Lake B, Aeppli G, Clausen K N, McMorrow D F, Lefmann K, Hussey N E,
Mangkorntong N, Nohara M, Takagi H, Mason T E and Schroder A 2001
\emph{Science} {\bf 291}, 1759

\bibitem{Lake2}
Lake B, Ronnow H M, Christensen N B, Aeppli G, Lefmann K, McMorrow D F, Vorderwisch P,
Smeibidl P, Mangkarntong N, Sasagawa T, Nohara M, Takagi H and Mason T E 2002
\emph{Nature} (London) {\bf 415}, 299

\bibitem{Hoffman}
Hoffman J E, Hudson E W, Lang K M, Madhavan V, Eisaki H, Uchida S and Davis J C
2002 \emph{Science} {\bf 295}, 466

\bibitem{Xue}
Song C -L, Wang Y -L, Cheng P, Jiang Y -P, Li W, Zhang T, Li Z, He K, Wang L -L,
Jia J -F, Hung H -H, Wu C -J, Ma X -C, Chen X and Xue Q -K
2011 \emph{Science} {\bf 332}, 1410

\bibitem{Hertz}
Hertz J 1976 \emph{Phys. Rev. B} {\bf 14}, 1165

\bibitem{Millis}
Millis A J 1993 \emph{Phys. Rev. B} {\bf 48}, 7183

\bibitem{Moriya}
Moriya T 1995 \emph{Spin Fluctuations in Itinerant Electron Magnetism}
(Springer-Verlag, Berlin, New York).

\bibitem{Belitz}
Belitz D, Kirkpatrick T R and Vojta T 1997 \emph{Phys. Rev. B} {\bf 55}, 9452

\bibitem{Rech}
Chubukov A V, P\'{e}pin C and Rech J  2004 \emph{Phys. Rev. Lett.} {\bf
92}, 147003; Rech J, P\'{e}pin C and Chubukov A V
2006 \emph{Phys. Rev. B} {\bf 74}, 195126

\bibitem{Abanov}
Abanov A and Chubukov A V 2004 \emph{Phys. Rev. Lett.} {\bf 93}, 255702

\bibitem{Metzner}
Strack P, Takei S and Metzner W 2010 \emph{Phys. Rev. B} {\bf 81}, 125103;
Thier S C and Metzner W 2011 \emph{Phys. Rev. B} {\bf 84}, 155133

\bibitem{Liu}
Liu G -Z, Wang J -R and Wang J 2012 \emph{Phys. Rev. B} {\bf 85}, 174525

\bibitem{Ando}
Ando Y, Segawa K, Komiya S and Lavrov A N 2002 \emph{Phys. Rev. Lett.}
{\bf 88}, 137005

\bibitem{Hinkov}
Hinkov V, Haug D, Fauque B, Bourges P, Sidis Y, Ivanov A, Bernhard C,
Lin C T and Keimer B 2008 \emph{Science} {\bf 319}, 597

\bibitem{Daou}
Daou R, Chang J, LeBoeuf D, Cyr-Choiniere O, Laliberte F, Doiron-Leyraud N,
Ramshaw B J, Liang R, Bonn D A, Hardy W N and Taillefer L 2010
\emph{Nature} (London) {\bf 463}, 519

\bibitem{Lawler}
Lawler M J, Fujita K, Lee Jhinhwan, Schmidt A R, Kohsaka Y,
Kim Ch K, Eisaki H, Uchida S, Davis J C, Sethna J P and
Kim E -A 2010 \emph{Nature} {\bf 466}, 347

\bibitem{Kivelson}
Kivelson S A, Fradkin E and Emery V J 1998 \emph{Nature} (London), {\bf
393}, 550

\bibitem{Metznernematic}
Halboth C J and Metzner W 2000 \emph{Phys. Rev. Lett.} {\bf 85}, 5162;
Metzner W, Rohe D and Andergassen S 2003 \emph{Phys. Rev. Lett.}
{\bf 91}, 066402; Dell'Anna L and Metzner W 2006 \emph{Phys. Rev. B}
{\bf 73}, 045127

\bibitem{Oganesyan}
Oganesyan V, Kivelson S A and Fradkin E 2001 \emph{Phys. Rev. B} {\bf 64},
195109

\bibitem{VojtaSachdev}
Vojta M, Zhang Y and Sachdev S 2000 \emph{Phys. Rev. B} {\bf 62}, 6721;
2000 \emph{Phys. Rev. Lett.} {\bf 85}, 4940

\bibitem{Kim}
Kim E -A, Lawler M J, Oreto P, Sachdev S, Fradkin E and Kivelson S A
2008 \emph{Phys. Rev. B} {\bf 77}, 184514

\bibitem{Huh}
Huh Y and Sachdev S 2008 \emph{Phys. Rev. B} {\bf 78}, 064512

\bibitem{Sachdevbook}
Sachdev S 2011 \emph{Quantum Phase Transitions}, Chap. 17 (Cambridge
University Press).

\bibitem{Xu}
Xu C, Qi Y and Sachdev S 2008 \emph{Phys. Rev. B} {\bf 78}, 134507

\bibitem{Fritz}
Fritz L and Sachdev S 2009 \emph{Phys. Rev. B} {\bf 80}, 144503

\bibitem{Lawler2}
Kim E -A  and Lawler M J 2010 \emph{Phys. Rev. B} {\bf 81}, 132501

\bibitem{WLK}
Wang J, Liu G -Z and Kleinert H 2011 \emph{Phys. Rev. B} {\bf 83}, 214503

\bibitem{Shankar94}
Shankar R 1994 \emph{Rev. Mod. Phys.} {\bf 66}, 129

\bibitem{Orenstein00}
Orenstein J and Millis A J 2000 \emph{Science} {\bf 288}, 468

\bibitem{Chiao}
Chiao M, Hill R W, Lupien Ch, Taillefer L, Lambert P, Gagnon R and Fournier P
2000 \emph{Phys. Rev. B} {\bf 62}, 3554

\bibitem{Lee93}
Lee P A 1993 \emph{Phys. Rev. Lett.} {\bf 71}, 1887

\bibitem{Durst}
Durst A and Lee P A 2000 \emph{Phys. Rev. B} {\bf 62}, 1270

\bibitem{Lee97}
Lee P A and Wen X -G 1997 \emph{Phys. Rev. Lett.} {\bf 78}, 4111;
Paramekanti A and Randeria M 2002 \emph{Phys. Rev. B} {\bf 66}, 214517


%\bibitem{Sachdev1999book}
%S. Sachdev, \emph{Quantum phase transitions} (Cambridge University
%Press, Cambridge, 1999)

%\bibitem{Halperin}
%B. I. Halperin, T. C. Lubensky, and S.-K. Ma, Phys. Rev. Lett. {\bf
%32}, 292 (1974).

%\bibitem{Fisher}
%E. Domany, D. Mukamel, and M. E. Fisher, Phys. Rev. B {\bf15}, 5432
%(1977); J. Rudnick, Phys. Rev. B {\bf18}, 1406 (1978); J.-H. Chen
%and T. C. Lubensky, Phys. Rev. B {\bf17}, 4274 (1978).

%\bibitem{Cardy1996}
%J. Cardy, \emph{Scaling and Renormalization in Statistical Physics}
%(Combridge University Press, Combridge, UK, 1996).



\end{thebibliography}
\end{document}